\documentclass[reprint,aps,pra,showpacs]{revtex4-1}
\usepackage{amsmath}
\usepackage{bm}
\usepackage{graphicx}
\usepackage{wrapfig}
\usepackage{epstopdf}
\usepackage{epsfig}
\usepackage{calrsfs}
\usepackage{IEEEtrantools}
\usepackage{grffile}

\newcommand{\la}{\langle}
\newcommand{\ra}{\rangle}
\newcommand{\nn}{\nonumber}

\newcommand{\om}{\omega}

\newcommand{\p}{\partial}
\newcommand{\Sch}{Schr\"{o}dinger}

\newcommand{\tht}{\theta}
\newcommand{\Omg}{\Omega}

\newcommand{\hb}{\hbar}

\begin{document}



\title{Analytic solution and pulse area theorem for three--level atoms}
\author{Gavriil Shchedrin$^{1,2}$, Chris O'Brien$^{1}$, Yuri Rostovtsev$^{3}$, and Marlan O. Scully$^{1,4,5}$}
\affiliation{
$^{1}$Texas A{\&}M University, College Station, TX, 77843, USA\\
$^{2}${Colorado School of Mines, Golden, COL, 80401, USA}\\
$^{3}$University of North Texas,	Denton, TX, 76203, USA\\
$^{4}$Princeton University, Princeton, NJ, 08544, USA\\
$^{5}$Baylor University, Waco, TX, 76798, USA
}

\begin{abstract}
We report an analytic solution for a three--level atom driven by arbitrary time-dependent electromagnetic pulses. In particular, we consider far--detuned driving pulses and show an excellent match between our analytic result and the numerical simulations. We use our solution to derive a pulse area theorem for three--level $V$ and $\Lambda$ systems without making the rotating wave approximation. 
Formulated as an energy conservation law, this pulse area theorem provides a simple picture for a pulse propagation through a three--level media.
\end{abstract}

\maketitle

\section{Introduction}
The classic McCall-Hahn theorem  \cite{Hahn_1969} shows that a soliton with an even-$\pi$ pulse area propagates freely through a strongly interacting two--level atomic media. This fundamental result has generated much interest in studying pulse propagation in two-level systems \cite{Lamb_1979}, in particular, in attempting to find accurate analytic expressions for a driven two--level system beyond the rotating wave approximation (RWA) \cite{swain1, schleich1,chris1}, 
arbitrary pulse propagation through a two-level media \cite{Yuri1, Yuri2, sarma1}, and control of a soliton propagation \cite{Boyd_2001, Shin_1999}.

The success in two-level media encouraged consideration of pulse propagation in three level media where new physical phenomena such as electromagnetically induced transparency in the $\Lambda$ scheme \cite{Harris_EIT, Olga_EIT, Lukin_2000} 
and lasing without inversion in the $V$ scheme \cite{Olga_LWI, Harris_LWI, Scully_LWI} were discovered. 
Despite the advances in three level media, a theory of soliton propagation and in particular a 
pulse area theorem for three level systems is currently missing. 
So far, soliton propagation in three level media has only been considered numerically \cite{vemuri1} or analytically treated for limited special cases \cite{Shin_1998, eberly1, bishop1, milonni1}.  

In this paper we explore the dynamics of three-level systems driven by two arbitrary time-dependent electromagnetic fields, without using the RWA. 
First we introduce a method based on the time evolution operator, which allows us to derive an analytic solution for three--level systems. 
Then we compare our analytic results with numerical simulations, to establish when our approach is useful. In particular, we show that the obtained solution works well for far-detuned pulses, when the ratio of the peak Rabi frequency to the detuning of central field frequency can be regarded as a small parameter.
Based on this solution we derive a pulse area theorem for three--level systems.
In the special case of a single resonant  field treated within the RWA, the pulse area theorem reduces to the classic McCall-Hahn theorem.

The pulse area theorem was motivated by the recent development of the new type of laser, 
based on quantum amplification by superradiant emission of radiation (QASER).
\cite{scully_q1, scully_q2}.   The two-level atomic media in the QASER description is driven by a strong far-off resonant field where the RWA breaks down. Therefore the classic McHall-Hahn theorem cannot be applied for this case. We have successfully applied the pulse area theorem that is valid beyond the RWA and achieved the parametric excitation of a driven two--level atomic media \cite{scully_q3}.

\section{Time evolution operator for driven three--level atoms}
\begin{figure}[t!]\label{f1}
\begin{center}
\includegraphics[angle=0, width=1.0\columnwidth]
{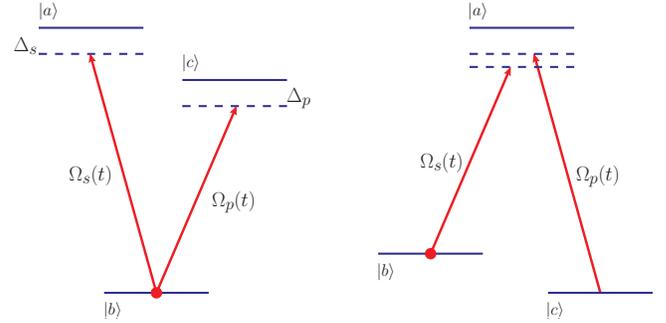}
\end{center}
 \caption{(Color online) The three level $V$ and $\Lambda$--systems interacting with the electromagnetic pulses $\Omega_{s}(t) $ and $\Omega_{p}(t)$. The atomic transition frequencies are 
 $\om_{ab}=\om_{a}-\om_{b}$ 
and $\om_{cb}=\om_{c}-\om_{b}$. The initial state is prepared in the ground state, $|\psi(t_{0})\rangle=|b\rangle$.}
\end{figure}

The exact Hamiltonian of a driven three-level $V$ system in the interaction picture is
\begin{equation}
H_{int} ^{V} (t)=-\hbar
\left(
\begin{array}{ccc}
 0 &  \Omega _s(t) e^{i\om_{ab} t} &0\\
 \Omega^{*} _s(t)e^{-i\om_{ab} t}  & 0 &   \Omega^{*} _{p}(t)e^{-i\om_{cb} t} \\
 0 &   \Omega _{p}(t)e^{i\om_{cb} t} & 0
\end{array}
\right).
\end{equation}
Here the time-dependent amplitudes, 
$\Omega_{s}(t) =\Omega_{s} \Sigma_{s} (t) \cos( \nu_{s} t)$ and  
$\Omega_{p}(t) =\Omega_{p} \Sigma_{p} (t) \cos( \nu_{p} t)$, 
are defined for arbitrary time--dependent  electromagnetic pulse envelopes $\Sigma_{s} (t)$ and $\Sigma_{p} (t)$.  The peak Rabi frequencies $\Omega_{s} = \wp_{ab} E_{s} / \hbar $  and  
$\Omega_{p} = \wp_{cb} E_{p} / \hbar $ are defined in terms of the maximal values of the electromagnetic fields $E_{s}$ and $E_{p}$ that drive the population between the ground state $|b\rangle$ and the excited states $|a\rangle$ and $|c\rangle$. The carrier driving frequencies are $\nu_{s}$ and $\nu_{p}$. The matrix elements of the dipole moments are $\wp_{ab}=e\la{a}|r|{b}\ra$ and  $\wp_{cb}=e\la{c}|r|{b}\ra$. 
The relative atomic transition frequencies  $\om_{ab}=\om_{a}-\om_{b}$ 
and $\om_{cb}=\om_{c}-\om_{b}$ are defined in terms of the energies of the individual atomic levels $\om_{a}$, $\om_{b}$, and $\om_{c}$.

We start with the derivation of an analytical solution for a  three--level system 
interacting with an arbitrarily time-dependent electromagnetic field. 
Our approach is based on the time evolution operator $U(t)$
represented by the time-ordered $T$ exponent of the Hamiltonian 
for a driven three--level system. The time evolution operator $U(t)$ keeps both slowly and rapidly oscillating terms in the Hamiltonian and therefore leads to a solution valid for an arbitrary time-dependent electromagnetic field beyond the RWA.
The time ordering procedure can be accomplished by breaking the total time interval $(t_{f}-t_{i})$ into $N$ infinitesimal intervals $ \Delta \tau = (t_{f}-t_{i})/ N $ and for each discrete time, 
$ t_{k}=t_{i}+\left[(2k-1)/2 \right] \Delta \tau ,$
we evaluate the original Hamiltonian $H_{int}(t_{k})$.
Then the time-evolution operator is represented by
\begin{align}
U(t) &=\widehat{T} \exp \left[ -\frac{i}{\hbar}\int_{t_{i}}^{t_{f}} dt' H_{int}(t') \right] = \nn \\
&\exp \left[ -\frac{i}{\hbar} \Delta \tau H_{int}(t_{N}) \right]\times \cdots \times\exp \left[ -\frac{i}{\hbar} \Delta \tau H_{int}(t_{1}) \right].
\end{align}
By collecting the infinitesimal contributions 
in the limit of an infinitely small time interval
$ \Delta \tau \longrightarrow 0$ and infinitely large
number of steps $N \longrightarrow \infty$
we arrive at the Magnus expansion \cite{Magnus1},
\begin{equation}
U(t)=\exp \left[ \sum_{n=1}^{\infty}S^{(n)} \right].
\end{equation}
The first few terms in the Magnus expansion can be obtained by means of the Baker--Campbell--Hausdorff formula
\begin{equation}
S^{(1)}(t)= -\frac{i}{\hbar} \int_{0}^{t} dt_{1} H_{int}(t_{1}), \label{Magnus1}
\end{equation}
\begin{equation}
S^{(2)}(t)= \left( -\frac{i}{\hbar} \right)^{2} \frac{1}{2} \int_{0}^{t} dt_{1} \int_{0}^{t_{1}} dt_{2} [H_{int}(t_{1}),H_{int}(t_{2})]. \label{Magnus2}
\end{equation}
This procedure can be further generalized to get an explicit expression for the $n${th} term in the Magnus expansion $S^{(n)}(t)$ in terms of integrals over sums of nested commutators \cite{Blanes1}.

Now we introduce our only approximation for a far-detuned field.
In this case we can perform perturbation theory by truncating the infinite series in the exponential keeping solely the leading order term in the Magnus expansion, given by Eq (\ref{Magnus1}). 
Here we will first focus on the $V$ scheme, and later apply the same method to the $\Lambda$ scheme. 
By projecting the time-evolution operator onto the ground state,
\begin{eqnarray}\label{projection1}
|\psi(t)\rangle=U(t)|b\rangle,
\end{eqnarray}
we obtain the solution of a three-level $V$ scheme beyond the rotating wave approximation for a general time-dependent pulse shape.
In order to find the state given by Eq. (\ref{projection1}) we need to diagonalize the matrix $S^{(1)}(t)$  which is a simple task after rewriting it in terms of the complex pulse areas
\begin{eqnarray}\label{vars1}
\tht_{s}(t)=\int_{0}^{t}{dt}\;\Omg_{s}(t) \exp{[i\om_{ab} t]},
\end{eqnarray}
\begin{eqnarray}\label{varp1}
\tht_{p}(t)=\int_{0}^{t}{dt}\;\Omg_{p}(t)\exp{[i\om_{cb} t]}.
\end{eqnarray}
Here $\Omg_{s}(t)$ is the entire time dependence of the field including the arbitrary time-dependent pulse shape along with slow and fast oscillations. 
This matrix can be easily diagonalized using the standard method of eigenvalues and eigenvectors. 
Then the three eigenvalues of the matrix are zero, positive and negative effective pulse areas,
\begin{eqnarray}\label{area1}
\theta(t)=\sqrt{|\theta_{s}|^{2}+|\theta_{p}|^{2}}.
\end{eqnarray}
With the initial condition for the population prepared  in the ground state, the solution for a three-level $V$ scheme is,
\begin{IEEEeqnarray}{c}\label{sol1}
|\psi^{V} (t) \ra =
i \theta _s(t) \frac{ \sin \left[ {\theta(t)}\right] }{{\theta(t)}}
 |a\rangle+
 \cos \left[{\theta(t)}\right]
  |b\rangle+\nn\\
i \theta_{p} (t)
 \frac{\sin \left[{\theta(t)}\right]}{{\theta(t)}}
  |c\rangle.
\end{IEEEeqnarray}

The same approach can be applied to the $\Lambda$ system, described by the  exact Hamiltonian $H ^{\Lambda}_{int}(t)$ in the interaction picture,
\begin{equation}
H ^{\Lambda}_{int}(t)=-
\hbar
\left(
\begin{array}{ccc}
 0 & \Omega _s(t) e^{i\omega_{ab} t} &   \Omega _{p }(t)e^{i\omega_{ac} t}\\
 \Omega^{*} _s(t)e^{-i\omega_{ab} t}  & 0 & 0 \\
  \Omega^{*} _{p }(t)e^{-i\omega_{ac} t} & 0 & 0
\end{array}
\right).
\end{equation}
Following the same method as for the $V$ scheme, we obtain the solution for the $\Lambda$ scheme in terms of the complex field areas
\begin{align} \label{sol2}
|\psi^{\Lambda}(t)\ra &=
 \frac{i \theta_{s}(t) \sin \left[\theta(t)\right] }{\theta(t)} 
|a\rangle
 + \frac{\left(
\left|\theta _{p}(t)\right|^{2} +
\left|\theta _{s}(t)\right|^{2}
 \cos \left[\theta(t)\right]
 \right)} {|\theta(t)|^{2}}
 |b \rangle \nn\\
 & +\frac{\theta_{s}\theta_{p}^{*} }
 {|\theta(t)|^{2}}
 \left(\cos \left[\theta(t)\right]-1\right)
 |c\rangle,
\end{align}
where the pulse area $\theta_{s}$ is given by Eq. (\ref{vars1}), while the pulse area $\theta_{p}$ is now modulated by the atomic transition frequency $\om_{ac}=\omega_{a}-\omega_{c}$ for a $\Lambda$ scheme compared to $\om_{cb}$ in the $V$ scheme,
\begin{eqnarray}\label{varp2}
\tht_{p}(t)=\int_{0}^{t}{dt}\;\Omg_{p}(t)\exp{[i\om_{ac} t]}.
\end{eqnarray}

In the special case of a resonant continuous wave (CW) driving field treated within the RWA, i.e. $\Omg_{s}(t) = \Omg_{s}e^{-i\nu_{s}t}$ and  $\Omg_{p}(t) =  \Omg_{p}e^{-i\nu _p t}$, we can simplify the complex area variables into
\begin{eqnarray}\label{rwavar1}
\tht_{s}(t)=\frac{\Omg_{s} t}{2}, \\
\tht_{p}(t)=\frac{\Omg_{p} t}{2}.
\end{eqnarray}
Then from Eq. (\ref{sol1}), we immediately obtain the conventional RWA solution for a three-level $V$ system,
\begin{equation}\label{rwa1}
|\psi^{V}(t)\rangle
=
 \frac{i \Omega _s}{\Omega} \sin\left[\frac{\Omega {t}}{2}\right]
   |a\rangle
 +
\cos\left[\frac{ \Omega {t}}{2}\right]
   |b\rangle
 +
 \frac{i \Omega _{p } }{\Omega} \sin\left[\frac{\Omega {t}}{2}\right]
  |c\rangle,
\end{equation}
where the effective Rabi frequency is 
\begin{eqnarray}\label{effect1}
\Omega=\sqrt{\Omega_{s}^{2}+\Omega_{p}^{2}}.
\end{eqnarray}
Likewise for the $\Lambda$ scheme, the solution given by Eq. (\ref{sol2})  simplifies under the resonant CW driving field to the conventional RWA solution,
\begin{align}
|\psi^{\Lambda}(t)\ra &=
 \frac{i \Omega_{s}  }{\Omega} \sin \left[\frac{\Omega t }{2}\right]
|a\rangle
+  \frac{1} {\Omega^{2}}
 \left(
\Omega^{2}_{p}+ \Omega^{2}_{s}
 \cos \left[\frac{\Omega t }{2}\right]
 \right)
 |b \rangle \nn \\
& +   \frac{\Omega_{s}\Omega_{p} }
 {\Omega^{2}}
 \left(\cos\left[\frac{\Omega t }{2}\right]-1\right)
 |c\rangle,
\end{align}
where the effective Rabi frequency $\Omega$ is given by (\ref{effect1}).

\begin{figure*}[t!]
\includegraphics[angle=0, width=1.8\columnwidth]{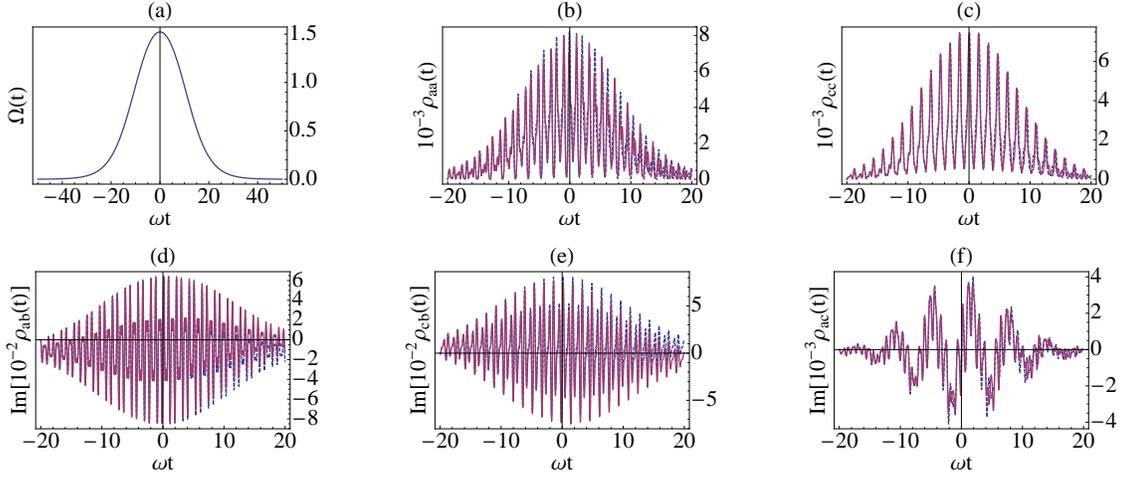}
\caption{(Color online) Atomic three-level V-system interacting with a far detuned electromagnetic continuous wave  pulses,  $\Omega_{s}(t)=\Omega_{s} \Sigma(t) \cos(\nu_{s} t)$ and $\Omega_{p}(t)=\Omega_{p} \Sigma(t) \cos(\nu_{p} t)$
 with $\nu_{s} = 3$, $\Omega_{s}=3/5$, and $\nu_{p} = 2$, $\Omega_{p}=1/2$.
 The pulse envelope is given by Eq. (\ref{pulse2}) with parameters $q=0$ and $\tau_p = 10$.
The driven atom has $\om_{ab}=12$ and $\om_{cb}=10$. Here 
 the dashed (blue) line is our approximate analytic solution from Eq. (\ref{sol1})
and  the solid (red) line is the numerical solution. For this case the two are almost perfectly matched.
Here we plot several key variables: 
(a) pulse envelope,
(b) population in level $a$,
(c) population in level $c$,
(d) imaginary part of the coherence $\rho _{ab}$,
(e) imaginary part of the coherence $\rho _{cb}$,
(f) imaginary part of the coherence $\rho _{ac}$.
}
\label{f2}
\end{figure*}
\begin{figure*}[t!]
\includegraphics[angle=0, width=1.8\columnwidth]{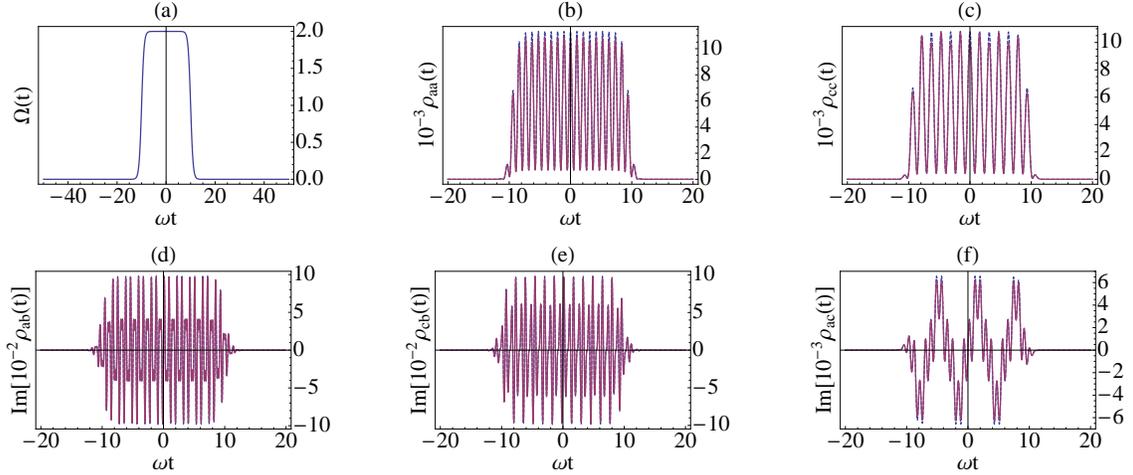}
\caption{(Color online) Same as Fig. (\ref{f2}) but for $q=1$ and $\tau_{p}=10$}
\label{f3}
\end{figure*}
%


\section{Analytic Solution vs. Numerical simlations}\label{numerics}

To compare our solution with the exact numerics we consider pulses of a general form
\begin{equation}
\Omega (t) = \Omega _R \Sigma (t) \cos( \nu t), \label{pulse1}
\end{equation}
where $\nu $ is the central frequency of the pulse, $\Omega_R = \wp_{ab} E_{0} / \hb $ is the peak pulse Rabi frequency, and
$\Sigma (t) $ is the pulse envelope.
Here we restrict our examples to pulses with the varying envelope function
\begin{equation}
\Sigma (t) = \tanh \left[ 10^{q} (t+ \tau_{p})/\tau_{p} \right] - 
\tanh \left[ 10^{q} (t-\tau_{p})/\tau_{p} \right], \label{pulse2}
\end{equation}
which gives pulse envelopes of duration $2\tau_p$ and with a shape controlled by the parameter $q$. For $q\geq{1}$ we essentially have a square pulse as shown in Fig.(\ref{f3}). 
With $q<1$ the square pulse smooths out until for $q  = 0 $ the envelope is nearly Gaussian  as can be seen in Fig.(\ref{f2}).

Now we can examine the approximation of dropping all but the first term of the Magnus expansion.
Considering the fields defined by Eq. (\ref{pulse1}) it can be seen that the $n$-th term in the Magnus expansion is proportional to the $n$-th power of the integral over the CW square pulse in the RWA
\begin{eqnarray}
\int^{t}_{0}{dt}\; \Omega(t)\exp{[i\om t]}\simeq{}\frac{\Omega_{R}}{\Delta}
\sin\left[\frac{\Delta t}{2} \right]
\exp[i \Delta t/2].
\end{eqnarray}
Thus the approximation of dropping higher order terms in the Magnus expansion of the evolution operator is essentially a perturbation theory in the evolution operator with the small parameter, ${\Omega_{R}}/{\Delta}\ll{1}$. Here $\Delta=\omega-\nu$ is the detuning between the central field frequency and the transition frequency (either $\omega _{ab}$ or $\omega_{cb}$).
To get higher accuracy one should keep the higher order terms given by Eq. (\ref{Magnus2}), but for the far-detuned case the leading term describes the dynamics of the light-matter interaction exceptionally well, as one can see from the agreement between the analytic solution and numerical simulations. 
For both adiabatic Gaussian-like pulse (as shown
in Fig. \ref{f2}) and non-adiabatic square pulse (as shown in Fig. \ref{f3})
there is a strong match between our analytic solution and the numerical simulations.

\section{Pulse area theorem for three-level atoms}

The obtained solution allows us to formulate the pulse area theorem for 
a driven three-level $V$ and $\Lambda$ systems. 
The Maxwell-{\Sch} equations in the slowly varying amplitude approximation for $V$ system are given by,
\begin{eqnarray}\label{max1}
\left( \frac{\partial }{\partial t}+c\frac{\partial }{\partial z}\right)
\frac{\partial }{\partial t} \theta_{s}(t)=
-i\Omg^{2}_{a}
\rho_{ab},
\end{eqnarray}
\begin{eqnarray}\label{max2}
\left( \frac{\partial }{\partial t}+c\frac{\partial }{\partial z}\right)
\frac{\partial }{\partial t} \theta_{p}(t)=
-i\Omg^{2}_{c}
\rho_{cb}
\end{eqnarray}
Here  $\Omg_{a}$, and $\Omg_{c}$ are the collective frequencies
proportional to the dipole moments associated with the transitions between 
the levels $a$ and $b$ and the levels $c$ and $b$ correspondingly,
\begin{align}
& \Omg_{a}^{2}=\frac{3}{8\pi}n\lambda_{ab}^{2}\gamma {c}, \\
&\Omg_{c}^{2}=\frac{3}{8\pi}n\lambda_{cb}^{2}\gamma {c} 
\end{align}
Here $n$ is the atomic density, $\lambda_{ab}$ and $\lambda_{cb}$
are the atomic wavelengths, and $\gamma$ is the spontaneous decay rates.

If we consider optically thin media then we can safely disregard space propagation in the Maxwell--{\Sch} equations (\ref{max1}) and (\ref{max2}). 
Combining the Maxwell--{\Sch} equations 
for $\theta_{s}(t)$ and $ \theta_{p}(t)$, multiplying them by 
$ {\p \theta^{*}_{s}}/{\partial t}$ and $ {\p \theta^{*}_{p}}/{\partial t}$, correspondingly, 
and adding the complex conjugates counterparts, the left hand side of Eqs. (\ref{max1}) and  (\ref{max2}) becomes a full-time derivative,
\begin{equation}
  \frac{\p}{\partial t}
  \left(
  \frac{1}{\Omega_{a}^{2}}
 \left|
    \frac{\p}{\partial t}
 \theta_{s}(t)
 \right|^{2}
+
  \frac{1}{\Omega_{c}^{2}}
 \left|
    \frac{\p}{\partial t}
 \theta_{p}(t)
 \right|^{2} 
 \right)
\end{equation}
Due to the density matrix equations the right hand sides of Eqs. (\ref{max1}) and (\ref{max2}) turn into full--time derivative of the population in the ground state,
\begin{align}
-i
\left[
\rho{}^{*}_{ab}(t)
  \frac{\p  \theta_{s}}{\partial t} -
\rho_{ab}(t) 
  \frac{\p \theta^{*}_{s}}{\partial t}
\right]
& \\\nn
-i\left[
\rho_{bc}(t)
  \frac{\p  \theta_{p}}{\partial t}
-
\rho{}^{*}_{bc}(t) 
  \frac{\p  \theta^{*}_{p}}{\partial t}
\right] & =
\frac{\p \rho_{bb}}{\p t}.
\end{align}
With the population in the ground state,
\begin{eqnarray}
\rho_{bb}(t)=\cos^{2}\left(\sqrt{|\theta_{s}|^{2}+|\theta_{p}|^{2}}\right)
\end{eqnarray}
we obtain the conservation law, 
\begin{align}
 \frac{\p}{\partial t}
 &
  \left(
  \frac{1}{\Omega_{a}^{2}}
 \left|
    \frac{\p}{\partial t}
 \theta_{s}(t)
 \right|^{2}
+
  \frac{1}{\Omega_{c}^{2}}
 \left|
    \frac{\p}{\partial t}
 \theta_{p}(t)
 \right|^{2} 
 \right)\\\nn=
 -\frac{\p}{\partial t} &\cos^{2}\left(\sqrt{|\theta_{s}|^{2}+|\theta_{p}|^{2}}\right)
\end{align}
Taking into account the initial conditions for both $\Omega_{s}$ and $\Omega_{p}$ pulses,
\begin{eqnarray}
 \left|
    \frac{\p}{\partial t}
 \theta(t_{0})
 \right|^{2}=|\theta(t_{0})|^{2}=0
\end{eqnarray}
we arrive at the pulse area theorem for three--level $V$ system in terms of the pulse areas 
of Eqs. (\ref{vars1}) and  (\ref{varp1}),
\begin{equation}
 \frac{1}{\Omega_{a}^{2}}
 \left|
    \frac{\p}{\partial t}
 \theta_{s}(t)
 \right|^{2}
+
  \frac{1}{\Omega_{c}^{2}}
 \left|
    \frac{\p}{\partial t}
 \theta_{p}(t)
 \right|^{2} 
=
\sin^{2}\left(\sqrt{|\theta_{s}|^{2}+|\theta_{p}|^{2}}\right)
\end{equation}
In terms of the energy densities of the fields we obtain,
\begin{equation}
 \left|
 \frac{\Omega_{s}(t)}{\Omega_{a}}
 \right|^{2}
+
 \left|
 \frac{\Omega_{p}(t)}{\Omega_{c}}
 \right|^{2}
=
\sin^{2}\left(\sqrt{|\theta_{s}|^{2}+|\theta_{p}|^{2}}\right)
\end{equation}
The pulse area theorem states that energy density of the fields $\Omega_{s}(t)$ and $\Omega_{p}(t)$ is transferred to the atomic population in the excited state of the system, which in turn is expressed in terms of the pulse areas $\theta_{s}(t)$ and $\theta_{p}(t)$.

The preceding procedure applied for a $\Lambda$ scheme yields the Pulse Area Theorem for a $\Lambda$--system, expressed in terms of the pulse areas Eqs. (\ref{vars1}) and (\ref{varp2}),
\begin{equation}
  \frac{1}{\Omega_{a}^{2}}
 \left|
    \frac{\p}{\partial t}
 \theta_{s}(t)
 \right|^{2}
+
  \frac{1}{\Omega_{c}^{2}}
 \left|
    \frac{\p}{\partial t}
 \theta_{p}(t)
 \right|^{2} 
=
 \frac{ |\theta_{s}|^{2} \sin^{2} \left(\sqrt{\left|\theta _s\right|^{2}+\left|\theta _{p }\right|^{2}}\right) }{\left|\theta _s\right|^{2}+\left|\theta _{p }\right|^{2}}.
\end{equation}

We see that the pulse evolution $ \theta_{s}(t)$ affects the evolution of the $\theta_{p}(t)$ and vice versa, through the modulation of the population in the excited state. Thus the pulse area theorem can be used to control and analyze pulse propagation through a three-level media.

To take into account the influence of decay mechanism and broadening phenomena one should solve the density matrix equations that includes the decay matrix $\Gamma$,
\begin{eqnarray}\label{densdec1}
\frac{\p}{\p t}\rho(t)=
-\frac{i}{\hbar}[H(t),\rho(t)]-\frac{1}{2}\{\Gamma,\rho\}
\end{eqnarray}
in terms of the commutator $[H(t),\rho(t)]$ and anticommutator $\{\Gamma,\rho\}$.
Now we introduce the effective non-Hermitian Hamiltonian 
\begin{eqnarray}\label{nonhermham1}
\mathcal{H}(t)=H(t)-\frac{i}{2}\hbar \Gamma
\end{eqnarray}
which is given in terms of the Hermitian Hamiltonian $H^{\dagger}=H$ and the decay matrix $\Gamma^{\dagger}=\Gamma$. Using the standard definition of the density matrix,
\begin{eqnarray}
\rho(t)=\sum_{}P_{\psi}
|\psi(t)\rangle
\langle\psi(t)|,
\end{eqnarray}
one can show that the {\Sch} equation with the effective non-Hermitian Hamiltonian  $\mathcal{H}(t)$,
\begin{eqnarray}
\frac{\p}{\p t}
|\psi(t)\rangle
=
-\frac{i}{\hbar}\mathcal{H}(t)
|\psi(t)\rangle
\end{eqnarray}
is equivalent to the density matrix equation (\ref{densdec1})  that includes the decay matrix $\Gamma$.
Therefore in order to take into account the decay mechanism and broadening phenomena
into the current approach based on the time evolution operator $U(t)$, one should consider the effective non-Hermitian Hamiltonian (\ref{nonhermham1}).

\section{Conclusion}

In conclusion we investigated the dynamics of three-level systems driven by two arbitrarily time-dependent electromagnetic fields, without referring to the rotating wave approximation (RWA). In particular, we considered far-detuned driving pulse envelopes where the rotating wave approximation breaks down and show excellent agreement between the analytic solution and the numerical simulations. The solution is based on the time evolution operator that operates with a time--dependent Hamiltonian and allows us to keep both slowly and rapidly oscillating terms.
With the solution we formulate the pulse area theorem for three--level $V$ system and $\Lambda$ systems without making the rotating wave approximation. 
Formulated as the energy conservation law, the pulse area theorem for three--level systems provides a tool for pulse manipulation and  coherent control of three--level atoms.

\section{Acknowledgments}
The authors gratefully acknowledge stimulating discussions with 
Philip Hemmer and David Lee. The research was supported by
the National Science Foundation Grants PHY-1241032 (INSPIRE CREATIV), PHY-1068554, EEC-0540832 (MIRTHE ERC) and the Robert A. Welch Foundation Award A-1261.

\bibliographystyle{apsrev4-1}
\bibliography{vscheme_bib3}

\begin{thebibliography}{26}%
\makeatletter
\providecommand \@ifxundefined [1]{%
 \@ifx{#1\undefined}
}%
\providecommand \@ifnum [1]{%
 \ifnum #1\expandafter \@firstoftwo
 \else \expandafter \@secondoftwo
 \fi
}%
\providecommand \@ifx [1]{%
 \ifx #1\expandafter \@firstoftwo
 \else \expandafter \@secondoftwo
 \fi
}%
\providecommand \natexlab [1]{#1}%
\providecommand \enquote  [1]{``#1''}%
\providecommand \bibnamefont  [1]{#1}%
\providecommand \bibfnamefont [1]{#1}%
\providecommand \citenamefont [1]{#1}%
\providecommand \href@noop [0]{\@secondoftwo}%
\providecommand \href [0]{\begingroup \@sanitize@url \@href}%
\providecommand \@href[1]{\@@startlink{#1}\@@href}%
\providecommand \@@href[1]{\endgroup#1\@@endlink}%
\providecommand \@sanitize@url [0]{\catcode `\\12\catcode `\$12\catcode
  `\&12\catcode `\#12\catcode `\^12\catcode `\_12\catcode `\%12\relax}%
\providecommand \@@startlink[1]{}%
\providecommand \@@endlink[0]{}%
\providecommand \url  [0]{\begingroup\@sanitize@url \@url }%
\providecommand \@url [1]{\endgroup\@href {#1}{\urlprefix }}%
\providecommand \urlprefix  [0]{URL }%
\providecommand \Eprint [0]{\href }%
\providecommand \doibase [0]{http://dx.doi.org/}%
\providecommand \selectlanguage [0]{\@gobble}%
\providecommand \bibinfo  [0]{\@secondoftwo}%
\providecommand \bibfield  [0]{\@secondoftwo}%
\providecommand \translation [1]{[#1]}%
\providecommand \BibitemOpen [0]{}%
\providecommand \bibitemStop [0]{}%
\providecommand \bibitemNoStop [0]{.\EOS\space}%
\providecommand \EOS [0]{\spacefactor3000\relax}%
\providecommand \BibitemShut  [1]{\csname bibitem#1\endcsname}%
\let\auto@bib@innerbib\@empty
\bibitem [{\citenamefont {McCall}\ and\ \citenamefont
  {Hahn}(1969)}]{Hahn_1969}%
  \BibitemOpen
  \bibfield  {author} {\bibinfo {author} {\bibfnamefont {S.~L.}\ \bibnamefont
  {McCall}}\ and\ \bibinfo {author} {\bibfnamefont {E.~L.}\ \bibnamefont
  {Hahn}},\ }\href@noop {} {\bibfield  {journal} {\bibinfo  {journal} {Phys.
  Rev.}\ }\textbf {\bibinfo {volume} {183}},\ \bibinfo {pages} {457} (\bibinfo
  {year} {1969})}\BibitemShut {NoStop}%
\bibitem [{\citenamefont {Lamb}(1979)}]{Lamb_1979}%
  \BibitemOpen
  \bibfield  {author} {\bibinfo {author} {\bibfnamefont {G.~L.}\ \bibnamefont
  {Lamb}},\ }\href@noop {} {\bibfield  {journal} {\bibinfo  {journal} {Rev.
  Mod. Phys.}\ }\textbf {\bibinfo {volume} {43}},\ \bibinfo {pages} {99}
  (\bibinfo {year} {1979})}\BibitemShut {NoStop}%
\bibitem [{\citenamefont {Swain}(1973)}]{swain1}%
  \BibitemOpen
  \bibfield  {author} {\bibinfo {author} {\bibfnamefont {S.}~\bibnamefont
  {Swain}},\ }\href@noop {} {\bibfield  {journal} {\bibinfo  {journal} {J.
  Phys. A: Math., Nucl. Gen.}\ }\textbf {\bibinfo {volume} {6}},\ \bibinfo
  {pages} {192} (\bibinfo {year} {1973})}\BibitemShut {NoStop}%
\bibitem [{\citenamefont {Giese}\ \emph {et~al.}(2013)\citenamefont {Giese},
  \citenamefont {Roura}, \citenamefont {Tackmann}, \citenamefont {Rasel},\ and\
  \citenamefont {Schleich}}]{schleich1}%
  \BibitemOpen
  \bibfield  {author} {\bibinfo {author} {\bibfnamefont {E.}~\bibnamefont
  {Giese}}, \bibinfo {author} {\bibfnamefont {A.}~\bibnamefont {Roura}},
  \bibinfo {author} {\bibfnamefont {G.}~\bibnamefont {Tackmann}}, \bibinfo
  {author} {\bibfnamefont {E.~M.}\ \bibnamefont {Rasel}}, \ and\ \bibinfo
  {author} {\bibfnamefont {W.~P.}\ \bibnamefont {Schleich}},\ }\href@noop {}
  {\bibfield  {journal} {\bibinfo  {journal} {Phys. Rev. A}\ }\textbf {\bibinfo
  {volume} {88}},\ \bibinfo {pages} {053608} (\bibinfo {year}
  {2013})}\BibitemShut {NoStop}%
\bibitem [{\citenamefont {Brien}\ and\ \citenamefont {Scully}(2015)}]{chris1}%
  \BibitemOpen
  \bibfield  {author} {\bibinfo {author} {\bibfnamefont {C.~O.}\ \bibnamefont
  {Brien}}\ and\ \bibinfo {author} {\bibfnamefont {M.~O.}\ \bibnamefont
  {Scully}},\ }\href@noop {} {\bibfield  {journal} {\bibinfo  {journal} {J.
  Mod. Optics}\ } (\bibinfo {year} {2015})}\BibitemShut {NoStop}%
\bibitem [{\citenamefont {Jha}\ and\ \citenamefont {Rostovtsev}(2010)}]{Yuri1}%
  \BibitemOpen
  \bibfield  {author} {\bibinfo {author} {\bibfnamefont {P.-K.}\ \bibnamefont
  {Jha}}\ and\ \bibinfo {author} {\bibfnamefont {Y.-V.}\ \bibnamefont
  {Rostovtsev}},\ }\href@noop {} {\bibfield  {journal} {\bibinfo  {journal}
  {Phys. Rev. A}\ }\textbf {\bibinfo {volume} {81}},\ \bibinfo {pages} {033827}
  (\bibinfo {year} {2010})}\BibitemShut {NoStop}%
\bibitem [{\citenamefont {Rostovtsev}\ and\ \citenamefont
  {Eleuch}(2010)}]{Yuri2}%
  \BibitemOpen
  \bibfield  {author} {\bibinfo {author} {\bibfnamefont {Y.-V.}\ \bibnamefont
  {Rostovtsev}}\ and\ \bibinfo {author} {\bibfnamefont {H.}~\bibnamefont
  {Eleuch}},\ }\href@noop {} {\bibfield  {journal} {\bibinfo  {journal} {J.
  Mod. Opt.}\ }\textbf {\bibinfo {volume} {57}},\ \bibinfo {pages} {1882}
  (\bibinfo {year} {2010})}\BibitemShut {NoStop}%
\bibitem [{\citenamefont {Barnes}\ and\ \citenamefont
  {Das~Sarma}(2012)}]{sarma1}%
  \BibitemOpen
  \bibfield  {author} {\bibinfo {author} {\bibfnamefont {E.}~\bibnamefont
  {Barnes}}\ and\ \bibinfo {author} {\bibfnamefont {S.}~\bibnamefont
  {Das~Sarma}},\ }\href {\doibase 10.1103/PhysRevLett.109.060401} {\bibfield
  {journal} {\bibinfo  {journal} {Phys. Rev. Lett.}\ }\textbf {\bibinfo
  {volume} {109}},\ \bibinfo {pages} {060401} (\bibinfo {year}
  {2012})}\BibitemShut {NoStop}%
\bibitem [{\citenamefont {Park}\ and\ \citenamefont {Boyd}(2001)}]{Boyd_2001}%
  \BibitemOpen
  \bibfield  {author} {\bibinfo {author} {\bibfnamefont {Q.-H.}\ \bibnamefont
  {Park}}\ and\ \bibinfo {author} {\bibfnamefont {R.~W.}\ \bibnamefont
  {Boyd}},\ }\href {\doibase 10.1103/PhysRevLett.86.2774} {\bibfield  {journal}
  {\bibinfo  {journal} {Phys. Rev. Lett.}\ }\textbf {\bibinfo {volume} {86}},\
  \bibinfo {pages} {2774} (\bibinfo {year} {2001})}\BibitemShut {NoStop}%
\bibitem [{\citenamefont {Park}\ and\ \citenamefont {Shin}(1999)}]{Shin_1999}%
  \BibitemOpen
  \bibfield  {author} {\bibinfo {author} {\bibfnamefont {Q.-H.}\ \bibnamefont
  {Park}}\ and\ \bibinfo {author} {\bibfnamefont {H.~J.}\ \bibnamefont
  {Shin}},\ }\href {\doibase 10.1103/PhysRevLett.82.4432} {\bibfield  {journal}
  {\bibinfo  {journal} {Phys. Rev. Lett.}\ }\textbf {\bibinfo {volume} {82}},\
  \bibinfo {pages} {4432} (\bibinfo {year} {1999})}\BibitemShut {NoStop}%
\bibitem [{\citenamefont {Harris}(1997)}]{Harris_EIT}%
  \BibitemOpen
  \bibfield  {author} {\bibinfo {author} {\bibfnamefont {S.~E.}\ \bibnamefont
  {Harris}},\ }\href@noop {} {\bibfield  {journal} {\bibinfo  {journal} {Phys.
  Today}\ }\textbf {\bibinfo {volume} {50}},\ \bibinfo {pages} {36} (\bibinfo
  {year} {1997})}\BibitemShut {NoStop}%
\bibitem [{\citenamefont {Kocharovskaya}\ and\ \citenamefont
  {Khanin}(1986)}]{Olga_EIT}%
  \BibitemOpen
  \bibfield  {author} {\bibinfo {author} {\bibfnamefont {O.~A.}\ \bibnamefont
  {Kocharovskaya}}\ and\ \bibinfo {author} {\bibfnamefont {Y.~I.}\ \bibnamefont
  {Khanin}},\ }\href@noop {} {\bibfield  {journal} {\bibinfo  {journal} {JETP}\
  }\textbf {\bibinfo {volume} {90}},\ \bibinfo {pages} {1610} (\bibinfo {year}
  {1986})}\BibitemShut {NoStop}%
\bibitem [{\citenamefont {Fleischhauer}\ and\ \citenamefont
  {Lukin}(2000)}]{Lukin_2000}%
  \BibitemOpen
  \bibfield  {author} {\bibinfo {author} {\bibfnamefont {M.}~\bibnamefont
  {Fleischhauer}}\ and\ \bibinfo {author} {\bibfnamefont {M.~D.}\ \bibnamefont
  {Lukin}},\ }\href {\doibase 10.1103/PhysRevLett.84.5094} {\bibfield
  {journal} {\bibinfo  {journal} {Phys. Rev. Lett.}\ }\textbf {\bibinfo
  {volume} {84}},\ \bibinfo {pages} {5094} (\bibinfo {year}
  {2000})}\BibitemShut {NoStop}%
\bibitem [{\citenamefont {Kocharovskaya}\ and\ \citenamefont
  {Khanin}(1988)}]{Olga_LWI}%
  \BibitemOpen
  \bibfield  {author} {\bibinfo {author} {\bibfnamefont {O.~A.}\ \bibnamefont
  {Kocharovskaya}}\ and\ \bibinfo {author} {\bibfnamefont {Y.~I.}\ \bibnamefont
  {Khanin}},\ }\href@noop {} {\bibfield  {journal} {\bibinfo  {journal} {JETP
  Lett.}\ }\textbf {\bibinfo {volume} {48}},\ \bibinfo {pages} {630} (\bibinfo
  {year} {1988})}\BibitemShut {NoStop}%
\bibitem [{\citenamefont {Harris}(1989)}]{Harris_LWI}%
  \BibitemOpen
  \bibfield  {author} {\bibinfo {author} {\bibfnamefont {S.~E.}\ \bibnamefont
  {Harris}},\ }\href@noop {} {\bibfield  {journal} {\bibinfo  {journal} {Phys.
  Rev. Lett.}\ }\textbf {\bibinfo {volume} {62}},\ \bibinfo {pages} {1033}
  (\bibinfo {year} {1989})}\BibitemShut {NoStop}%
\bibitem [{\citenamefont {Scully}\ \emph {et~al.}(1989)\citenamefont {Scully},
  \citenamefont {Zhu},\ and\ \citenamefont {Gavrielides}}]{Scully_LWI}%
  \BibitemOpen
  \bibfield  {author} {\bibinfo {author} {\bibfnamefont {M.~O.}\ \bibnamefont
  {Scully}}, \bibinfo {author} {\bibfnamefont {S.-Y.}\ \bibnamefont {Zhu}}, \
  and\ \bibinfo {author} {\bibfnamefont {A.}~\bibnamefont {Gavrielides}},\
  }\href@noop {} {\bibfield  {journal} {\bibinfo  {journal} {Phys. Rev. Lett.}\
  }\textbf {\bibinfo {volume} {62}},\ \bibinfo {pages} {2813} (\bibinfo {year}
  {1989})}\BibitemShut {NoStop}%
\bibitem [{\citenamefont {Vemuri}\ \emph {et~al.}(1997)\citenamefont {Vemuri},
  \citenamefont {Agarwal},\ and\ \citenamefont {Vasavada}}]{vemuri1}%
  \BibitemOpen
  \bibfield  {author} {\bibinfo {author} {\bibfnamefont {G.}~\bibnamefont
  {Vemuri}}, \bibinfo {author} {\bibfnamefont {G.~S.}\ \bibnamefont {Agarwal}},
  \ and\ \bibinfo {author} {\bibfnamefont {K.~V.}\ \bibnamefont {Vasavada}},\
  }\href {\doibase 10.1103/PhysRevLett.79.3889} {\bibfield  {journal} {\bibinfo
   {journal} {Phys. Rev. Lett.}\ }\textbf {\bibinfo {volume} {79}},\ \bibinfo
  {pages} {3889} (\bibinfo {year} {1997})}\BibitemShut {NoStop}%
\bibitem [{\citenamefont {Park}\ and\ \citenamefont {Shin}(1998)}]{Shin_1998}%
  \BibitemOpen
  \bibfield  {author} {\bibinfo {author} {\bibfnamefont {Q.-H.}\ \bibnamefont
  {Park}}\ and\ \bibinfo {author} {\bibfnamefont {H.~J.}\ \bibnamefont
  {Shin}},\ }\href {\doibase 10.1103/PhysRevA.57.4643} {\bibfield  {journal}
  {\bibinfo  {journal} {Phys. Rev. A}\ }\textbf {\bibinfo {volume} {57}},\
  \bibinfo {pages} {4643} (\bibinfo {year} {1998})}\BibitemShut {NoStop}%
\bibitem [{\citenamefont {Eberly}\ and\ \citenamefont
  {Kozlov}(2002)}]{eberly1}%
  \BibitemOpen
  \bibfield  {author} {\bibinfo {author} {\bibfnamefont {J.~H.}\ \bibnamefont
  {Eberly}}\ and\ \bibinfo {author} {\bibfnamefont {V.~V.}\ \bibnamefont
  {Kozlov}},\ }\href {\doibase 10.1103/PhysRevLett.88.243604} {\bibfield
  {journal} {\bibinfo  {journal} {Phys. Rev. Lett.}\ }\textbf {\bibinfo
  {volume} {88}},\ \bibinfo {pages} {243604} (\bibinfo {year}
  {2002})}\BibitemShut {NoStop}%
\bibitem [{\citenamefont {Rybin}\ \emph {et~al.}(2005)\citenamefont {Rybin},
  \citenamefont {Vadeiko},\ and\ \citenamefont {Bishop}}]{bishop1}%
  \BibitemOpen
  \bibfield  {author} {\bibinfo {author} {\bibfnamefont {A.~V.}\ \bibnamefont
  {Rybin}}, \bibinfo {author} {\bibfnamefont {I.~P.}\ \bibnamefont {Vadeiko}},
  \ and\ \bibinfo {author} {\bibfnamefont {A.~R.}\ \bibnamefont {Bishop}},\
  }\href {\doibase 10.1103/PhysRevE.72.026613} {\bibfield  {journal} {\bibinfo
  {journal} {Phys. Rev. E}\ }\textbf {\bibinfo {volume} {72}},\ \bibinfo
  {pages} {026613} (\bibinfo {year} {2005})}\BibitemShut {NoStop}%
\bibitem [{\citenamefont {Ackerhalt}\ and\ \citenamefont
  {Milonni}(1986)}]{milonni1}%
  \BibitemOpen
  \bibfield  {author} {\bibinfo {author} {\bibfnamefont {J.~R.}\ \bibnamefont
  {Ackerhalt}}\ and\ \bibinfo {author} {\bibfnamefont {P.~W.}\ \bibnamefont
  {Milonni}},\ }\href {\doibase 10.1103/PhysRevA.33.3185} {\bibfield  {journal}
  {\bibinfo  {journal} {Phys. Rev. A}\ }\textbf {\bibinfo {volume} {33}},\
  \bibinfo {pages} {3185} (\bibinfo {year} {1986})}\BibitemShut {NoStop}%
\bibitem [{\citenamefont {Svidzinsky}\ \emph {et~al.}(2013)\citenamefont
  {Svidzinsky}, \citenamefont {Yuan},\ and\ \citenamefont
  {Scully}}]{scully_q1}%
  \BibitemOpen
  \bibfield  {author} {\bibinfo {author} {\bibfnamefont {A.~A.}\ \bibnamefont
  {Svidzinsky}}, \bibinfo {author} {\bibfnamefont {L.}~\bibnamefont {Yuan}}, \
  and\ \bibinfo {author} {\bibfnamefont {M.~O.}\ \bibnamefont {Scully}},\
  }\href {\doibase 10.1103/PhysRevX.3.041001} {\bibfield  {journal} {\bibinfo
  {journal} {Phys. Rev. X}\ }\textbf {\bibinfo {volume} {3}},\ \bibinfo {pages}
  {041001} (\bibinfo {year} {2013})}\BibitemShut {NoStop}%
\bibitem [{\citenamefont {Scully}(2014)}]{scully_q2}%
  \BibitemOpen
  \bibfield  {author} {\bibinfo {author} {\bibfnamefont {M.~O.}\ \bibnamefont
  {Scully}},\ }\href {http://stacks.iop.org/1555-6611/24/i=9/a=094014}
  {\bibfield  {journal} {\bibinfo  {journal} {Las. Phys.}\ }\textbf {\bibinfo
  {volume} {24}},\ \bibinfo {pages} {094014} (\bibinfo {year}
  {2014})}\BibitemShut {NoStop}%
\bibitem [{\citenamefont {Shchedrin}\ \emph {et~al.}(2015)\citenamefont
  {Shchedrin}, \citenamefont {Rostovtsev}, \citenamefont {Zhang},\ and\
  \citenamefont {Scully}}]{scully_q3}%
  \BibitemOpen
  \bibfield  {author} {\bibinfo {author} {\bibfnamefont {G.}~\bibnamefont
  {Shchedrin}}, \bibinfo {author} {\bibfnamefont {Y.}~\bibnamefont
  {Rostovtsev}}, \bibinfo {author} {\bibfnamefont {X.}~\bibnamefont {Zhang}}, \
  and\ \bibinfo {author} {\bibfnamefont {M.~O.}\ \bibnamefont {Scully}},\
  }\enquote {\bibinfo {title} {New approach to quantum amplification by
  superradiant emission of radiation},}\ \ (\bibinfo  {publisher} {World
  Scientific Publishing},\ \bibinfo {year} {2015})\ Chap.~\bibinfo {chapter}
  {9}\BibitemShut {NoStop}%
\bibitem [{\citenamefont {Magnus}(1954)}]{Magnus1}%
  \BibitemOpen
  \bibfield  {author} {\bibinfo {author} {\bibfnamefont {W.}~\bibnamefont
  {Magnus}},\ }\href@noop {} {\bibfield  {journal} {\bibinfo  {journal} {Comm.
  Pure Appl. Math.}\ }\textbf {\bibinfo {volume} {7}},\ \bibinfo {pages} {649}
  (\bibinfo {year} {1954})}\BibitemShut {NoStop}%
\bibitem [{\citenamefont {Blanes}\ \emph {et~al.}(2009)\citenamefont {Blanes},
  \citenamefont {Casas}, \citenamefont {Oteo},\ and\ \citenamefont
  {Ros}}]{Blanes1}%
  \BibitemOpen
  \bibfield  {author} {\bibinfo {author} {\bibfnamefont {S.}~\bibnamefont
  {Blanes}}, \bibinfo {author} {\bibfnamefont {F.}~\bibnamefont {Casas}},
  \bibinfo {author} {\bibfnamefont {J.}~\bibnamefont {Oteo}}, \ and\ \bibinfo
  {author} {\bibfnamefont {J.}~\bibnamefont {Ros}},\ }\href@noop {} {\bibfield
  {journal} {\bibinfo  {journal} {Phys. Rep.}\ }\textbf {\bibinfo {volume}
  {470}},\ \bibinfo {pages} {151} (\bibinfo {year} {2009})}\BibitemShut
  {NoStop}%
\end{thebibliography}%

\end{document}